# A new approach for determining optimal placement of PM$_{2.5}$ air quality sensors: case study for the contiguous United States

[1]Makoto M. Kelp, [2]Samuel Lin, [3]J. Nathan Kutz, and [4]Loretta J. Mickley

[1]Department of Earth and Planetary Sciences, Harvard University, Cambridge, MA 02138, USA
[2]Department of Computer Science, Harvard University, Cambridge, MA 02138, USA
[3]Department of Applied Mathematics, University of Washington, Seattle, WA 98195, USA
[4]John A. Paulson School of Engineering and Applied Sciences, Harvard University, Cambridge, MA 01238, USA

E-mail: mkelp@g.harvard.edu

## Abstract

Considerable financial resources are allocated for measuring ambient air pollution in the United States, yet the locations for these monitoring sites may not be optimized to capture the full extent of current pollution variability. Prior research on best sensor placement for monitoring fine particulate matter (PM$_{2.5}$) pollution is scarce: most studies do not span areas larger than a medium-sized city or examine timescales longer than one week. Here we present a pilot study using multiresolution modal decomposition (mrDMD) to identify the optimal placement of PM$_{2.5}$ sensors from 2000-2016 over the contiguous United States. This novel approach incorporates the variation of PM$_{2.5}$ on timescales ranging from one day to over a decade to capture air pollution variability. We find that the mrDMD algorithm identifies high-priority sensor locations in the western United States, but a significantly lower density of sensors than expected along the eastern coast, where a large number of EPA PM$_{2.5}$ monitors currently reside. Specifically, 53% of mrDMD optimized sensor locations are west of the 100$^{th}$ meridian, compared to only 32% in the current EPA network. The mrDMD sensor locations can capture PM$_{2.5}$ from wildfires and high pollution events, with particularly high skill in the West. These results suggest significant gaps in the current EPA monitoring network in the San Joaquin Valley in California, northern California, and in the Pacific Northwest (Idaho, and Eastern Washington and Oregon). Our framework diagnoses where to place air quality sensors so that they can best monitor smoke from wildfires. Our framework may also be applied to urban areas for equitable placement of PM$_{2.5}$ monitors.

**Introduction**

Air pollution consisting of fine particulate matter ($PM_{2.5}$) presents a major environmental risk to public health, causing 4.1 million premature deaths worldwide in 2019 (Murray et al., 2020). Given the financial cost and limited resources allocated for measuring ambient air pollution, however, optimizing sensor placement is a central mathematical challenge. Monitoring $PM_{2.5}$ is especially difficult as there are many outdoor sources including direct emissions from wildland and agricultural fires (Cusworth et al., 2018; Johnston et al., 2012), incomplete fuel combustion from vehicles and industrial processes (Bond et al., 2007), and secondary chemical formation from gas-phase precursors of anthropogenic (McDuffie et al., 2021) or biogenic (Goldstein et al., 2009) origin. Furthermore, the formation and transport of $PM_{2.5}$ are both sensitive to meteorological conditions from the local to synoptic scales (Tai et al., 2012), including, for example, convective events and horizontal winds that can disperse pollution as well as precipitation that can immediately remove particles. Here we provide an optimization framework that takes into account the multiscale variability of $PM_{2.5}$ to determine the placement of a minimum number of air pollution sensors for maximal spatiotemporal coverage.

In the United States, monitoring of air pollution has historically relied on the Environmental Protection Agency (EPA) network of sampling sites (US EPA, 2013). These monitoring locations are used to assess local and regional attainment of the National Ambient Air Quality Standard (NAAQS), to analyze air pollution impacts on public health, and to validate satellite measurements and air quality models. However, these sites are not equally distributed across the United States and may not adequately sample the full range of concentrations (Di et al., 2019). A greater number of sites are in the eastern United States, along the western coast, and in urban areas, with relatively fewer sites in mountainous regions and rural areas. Due in part to the high cost of maintenance, new monitoring sites are added infrequently (US EPA, 2020).

Improvements in low-cost sensor technologies coupled with the rise of citizen science air quality monitoring offers potential for greater spatial coverage of pollution observations, but even so these sensor locations may not provide adequate coverage. Many low-cost sensors can report measurements publicly in real-time (Kumar et al., 2015; Parmar et al., 2017; Snyder et al., 2013). These sensors, deployed by citizen science and crowdsourced organizations such as Purple Air (Barkjohn et al., 2021), represent a core component of an Internet of Things network for monitoring air pollution in cities (Dhingra et al., 2019; Toma et al., 2019). However, this promising democratization of air pollution monitoring may inadequately capture the spatial distribution and variability of $PM_{2.5}$. Crowdsourced datasets often rely on volunteers who are responsible for installation and upkeep of each sensor, resulting in deployment in predominantly white areas characterized by higher incomes and levels of education relative to the US census tracts with EPA monitors (deSouza and Kinney, 2021). Furthermore, areas with a greater density of low-cost sensors report lower annual-average $PM_{2.5}$ concentrations than the EPA monitors in all states expect California (deSouza and Kinney, 2021). To be sure, such citizen science efforts are well-intentioned, but they exacerbate disparities in the spatial coverage of $PM_{2.5}$ monitors and limit the pursuit of environmental justice (Sorensen et al., 2019).

Few studies have considered both spatial features and temporal dynamics of air pollution in determining the optimal placement of outdoor $PM_{2.5}$ sensors. While strategies based on computational fluid dynamics can locate optimal placement of indoor sensors (Löhner and Camelli, 2005; Waeytens and Sadr, 2018), such approaches are infeasible over large atmospheric domains. Mukherjee et al. (2020) developed an optimal sensor placement algorithm for mobile monitoring of $PM_{2.5}$ and $SO_2$ in Atlanta, Georgia, but such an algorithm can be applied for just a short time period (hours to a day). For a case study in Cambridge, U.K, Sun et al. (2019) created an algorithm that considers areas with $PM_{2.5}$ hotspots such as high-traffic roads and locations such as hospitals and elderly care homes where more vulnerable populations reside. However, this framework does not guarantee capturing representative $PM_{2.5}$ concentrations across a broader region. Here we build on prior studies that determined the optimal placement of sensors using modal decomposition techniques in the domains of ocean monitoring (Yildirim et al., 2009), fluid dynamics (Bai et al., 2017), and neuroscience (Brunton et al., 2016). Application of such techniques is yet nascent for air pollution, which is characterized by its own unique multiscale behavior. To our knowledge, there exist no studies of optimal air pollution sensor placement in the United States that span regions larger than an urban area or time scales longer than one week.

In this pilot study, we demonstrate a data-driven approach that determines the optimal placement of sensors to capture $PM_{2.5}$ concentrations and variability across the contiguous United States on time scales ranging from days to over a decade. We employ multiresolution dynamic mode decomposition (mrDMD), which recursively decomposes a dataset into low-rank spatial modes and their temporal Fourier dynamics (Kutz et al., 2016; Manohar et al., 2019). This algorithm allows us to create a library of modes that not only captures $PM_{2.5}$ concentrations spatially and temporally on short (daily) and long-term (years to decade) timescales, but also incorporates information from significant transient phenomena, such as wildfires and temperature inversions, that would otherwise be discarded or averaged out using similar data reduction techniques. We emphasize that the current distribution of over 2000 EPA monitors is needed to keep track of adherence to the NAAQS. However, the modeling approach and results presented here may be used to locate gaps in the EPA sensor network and inform locations of future $PM_{2.5}$ monitors.

**Data and Methods**

2.1 PM$_{2.5}$ dataset

We use a dataset consisting of daily PM$_{2.5}$ concentrations for the contiguous United States for the period January 2000 to December 2016. The dataset was produced through a data fusion method involving ensemble machine learning that combines surface monitoring measurements, satellite aerosol optical depth, land-use data, and chemical transport model results, among other variables (Di et al., 2021, 2019). The high spatial and temporal resolution of this PM$_{2.5}$ dataset reveals spatial variability across the United States, including a stark east-west gradient, with the eastern United States experiencing relatively higher PM$_{2.5}$ concentration than the West. Hotspots in the dataset reveal the wide diversity of PM$_{2.5}$ sources, such as emissions from power plants in the Midwest, organic aerosol produced from biogenic volatile organic compounds in the Southeast, and wildfires and associated smoke in the West. We coarsen the dataset's 1 km × 1 km resolution to 10 km × 10 km to yield a spatial grid of 871 × 413 cells. Missing values and erroneous observations are removed from the dataset by applying a mask to the spatial grid.

2.2 Dynamic mode decomposition (DMD)

To isolate the underlying multiscale variability of PM$_{2.5}$, we use a DMD modeling approach. The DMD method provides a spatiotemporal decomposition of data into a set of dynamic modes derived from snapshots or measurements of a given system in time. DMD is similar to principal components analysis (PCA) as they are both dimensionality reduction algorithms that take advantage of underlying low-rank features found in large datasets. The resulting modes of these methods reveal the dominant spatial patterns that account for variation in the dataset. However, unlike PCA, which does not model temporal dynamics of time-series data explicitly, DMD computes a set of modes with an exponential temporal dependence whose imaginary components model an oscillating time frequency characterizing the modal dynamics. Importantly, DMD can accommodate heterogenous spatial and temporal sampling. For noisy datasets, DMD can be ensembled in order to produce uncertainty metrics (Sashidhar and Kutz, 2021). Many variants of DMD exist (Askham and Kutz, 2017; Li et al., 2017; Noack et al., 2015; Sashidhar and Kutz, 2021; Tissot et al., 2014), but here we follow Manohar et al. (2018), as described in the supplement section 1. We apply the DMD method to the PM$_{2.5}$ dataset to construct a library of modes that reveals the dominant spatiotemporal patterns in the dataset.

2.3 Multiresolution dynamic mode decomposition (mrDMD)

Recently, a multiresolution extension to the DMD algorithm has been proposed (Kutz et al., 2016; Manohar et al., 2019), which allows for the incorporation of significant transient phenomenon into the DMD library. Much like a multiresolution analysis using wavelets (Kutz 2013), the mrDMD algorithm recursively removes low frequency, or slowly varying, signals to allow separation of background (i.e., average PM$_{2.5}$ concentrations) from foreground (i.e., high pollution episodes) in the PM$_{2.5}$ dataset on varying timescales specified by the user. The mrDMD algorithm is more precise in capturing spatiotemporal variability than methods based on singular value decomposition such as DMD and PCA (Manohar et al., 2019).

The number of snapshots *M* from our dataset are chosen so that the DMD modes provide a full rank approximation of the dynamics observed. Thus, *M* should be representative of all high- and low-frequency content of the dataset. The DMD algorithm is then applied recursively on the dataset. The slowest *m1* modes are removed in the initial pass, and remaining snapshots are divided into *M/2* segments and are fed into the DMD algorithm again. The slowest *m2* modes are removed, and this process is continued until a termination condition is reached. The mrDMD separates the DMD approximate solution in the first pass as follows:

Eq1)
$$X_{mrDMD}(t) = \sum_{k=1}^{M} b_k(0)\psi_k^{(1)}(\xi)\exp(\omega_k t)$$
$$= \underbrace{\sum_{k=1}^{m_1} b_k(0)\psi_k^{(1)}(\xi)\exp(\omega_k t)}_{(slow\ modes)} + \underbrace{\sum_{k=m_1+1}^{M} b_k(0)\psi_k^{(1)}(\xi)\exp(\omega_k t)}_{(fast\ modes)}$$

where $b_k(0)$ is the initial amplitude, $\xi$ is the spatial coordinates of each mode, $\psi_k^{(1)}$ represents the DMD modes (eigenvectors) obtained from the complete snapshot matrix, and $\exp(\omega_k t)$ are the corresponding eigenvalues. After the first level of decomposition, the fast modes are again decomposed into two matrices:

$$\text{Eq2)}\quad X_{M/2} = X_{M/2}^{(1)} + X_{M/2}^{(2)}$$

At this level of decomposition, the first matrix $X_{M/2}^{(1)}$ represents slower modes and $X_{M/2}^{(2)}$ represents faster modes, which undergo decomposition again. This process is repeated recursively by removing slower modes in each iteration. At each level of decomposition, the separation of the foreground mode from the background is deemed significant if its eigenvalues exceed a user-set tolerance. Here, we set this tolerance to a standard value of $1\times10^{-2}$.

A formal expansion of the mrDMD theory and modeling approach may be found elsewhere (Kutz et al., 2016; Manohar et al., 2019).

The mrDMD algorithm operates by decreasing the time domain by a factor of two at each successive decomposition level. We apply mrDMD to training windows starting at 11.4 years (*M*=4096 days), followed by 13 decomposition levels so that the shortest frequency is daily. This approach thus yields a long-term mode characterizing the average $PM_{2.5}$ concentrations over 11.4 years, with potential identification of transient pollution events spanning timescales from 5.7 years (2048 days) to one day. To analyze all $PM_{2.5}$ data from 2000 to 2016 employing these decomposition levels, we divide the dataset into two time-windows: January 2000 to March 2011 (first 4096 days) and September 2005 to December 2016 (last 4096 days).

2.4 Optimal sensor placement

We use the matrix libraries containing all DMD and mrDMD modes as tailored basis sets $\psi_r \in \mathbb{R}^{n \times r}$ to optimize for sensor placement. We identify the optimal sensor locations by employing QR pivoting to our DMD and mrDMD basis sets (Heck et al., 1998; Manohar et al., 2018). QR pivoting is a "greedy" selection algorithm that is computationally efficient for finding near-optimal sensor locations. Greedy approaches are often favored over other optimization techniques as the true optimal solution often involves a combinatorially intractable optimization.

QR column pivoting identifies rows in the modal library $\psi_r$ with the highest 2-norm, which corresponds to locations with the largest PM$_{2.5}$ modal frequencies and therefore greatest variability. The reduced matrix QR factorization with column pivoting decomposes a matrix $A \in \mathbb{R}^{m \times n}$ into a unitary matrix $Q$, an upper-triangular matrix $R$, and a column permutation matrix $C^T$ such that $AC^T = QR$. Thus, the QR factorization with column pivoting yields $r$ point sensors (pivots) that best sample the $r$ tailored basis modes $\psi_r$:

$$\text{Eq3)} \quad \psi_r^T C^T = QR$$

That is, each QR pivot identifies those spatial locations in the modal library that exhibit the most variability and where sensor placement captures significant pollution episodes above background concentrations.

A formal expansion of the sparse sensor placement approach may be found elsewhere (Brunton and Kutz, 2019; Kutz et al., 2016; Manohar et al., 2019, 2018).

In the supplement, we reconstruct the full field of PM$_{2.5}$ concentrations across the contiguous United States using measurements from only those sensor locations diagnosed as optimal in the step above. These reconstructions serve as validation of the DMD and mrDMD libraries and the ability of these algorithms to select sensor locations with high information content.

**Results and Discussion**

Here we present the mrDMD results and analysis. The modal library of DMD is less interpretable than that of mrDMD, and we present the DMD results and analysis in the supplement.

3.1 mrDMD modes describing $PM_{2.5}$ variability

The mrDMD modal library reveals that most significant air pollution episodes occur on time scales between one week and one month, with many of these events arising from wildfires. We diagnose these wildfire modes by visual inspection based on regions with known fire activity and minimal industrial $PM_{2.5}$ sources. Figure 1 presents the resulting mrDMD modal maps in the time-frequency domain. Figure 1a corresponds to the first 11.4 years of the time period (January 2000 to March 2011) and Figure 1b corresponds to the last 11.4 years (September 2005 to December 2016). The background mode in both training windows corresponds to the mean $PM_{2.5}$ concentration distribution for that window. Both background modes display a sharp east-west gradient with most of the eastern United States, except the Appalachian Mountains and some remote areas of Maine, having relatively higher $PM_{2.5}$ concentration than the West. This pattern agrees with the observed annual mean spatial distribution of $PM_{2.5}$ (Tai et al., 2010; Di et al., 2019). The background mode for the second half of the training period exhibits smaller modal amplitudes in hotspot regions than the first half. This trend is consistent with Di et al., who reported that $PM_{2.5}$ concentrations decreased noticeably after 2008 due to a combination of the economic recession and emission controls on coal-fired power plants. The 2000-2008 annual average $PM_{2.5}$ concentration across the contiguous United States was 7.8 $\mu g/m^3$; the 2009-2016 annual average was 6.2 $\mu g/m^3$.

The mrDMD method also identifies relatively short-term air pollution episodes on timescales of less than a year. Most episodes occur on timescales between one week and one month. However, we find an especially long-lived mode over Los Angeles and the San Joaquin Valley, California, lasting for 256 days from September 2007 to May 2008. This mode may correspond to intense winter $PM_{2.5}$ due to temperature inversions trapping emissions in the southern California basin. From 2007 to 2008, the San Joaquin Valley reported significantly higher $PM_{2.5}$ concentrations than are typical, with 66 days above the 24-h 35 $\mu g/m^3$ standard in 2008 (US EPA, 2021). The 2007 Murphy Complex Fire and other wildfires in Idaho are captured as a significant mode occurring on a time scale of 64 days from June to August. Although the lifetime of $PM_{2.5}$ in the atmosphere is ~one week and is sensitive to meteorology such as rainout, no pollution modes occur on timescales less than a week. We attribute the absence of modes on this timescale to the eigenvalue tolerance used in the mrDMD algorithm. If we tighten the tolerance, then more significant modes would be found at daily timescales, but such a tolerance would result in higher-frequency (i.e., noisy) and less-interpretable modal maps, making it more difficult to discern prolonged pollution events. Overall, we detect 72 significant pollution modes during the first time window and 77 during the second time window. We find 27 wildfire modes (38% of all modes) in the first window and 46 wildfire modes (60% of all modes) in the second window.

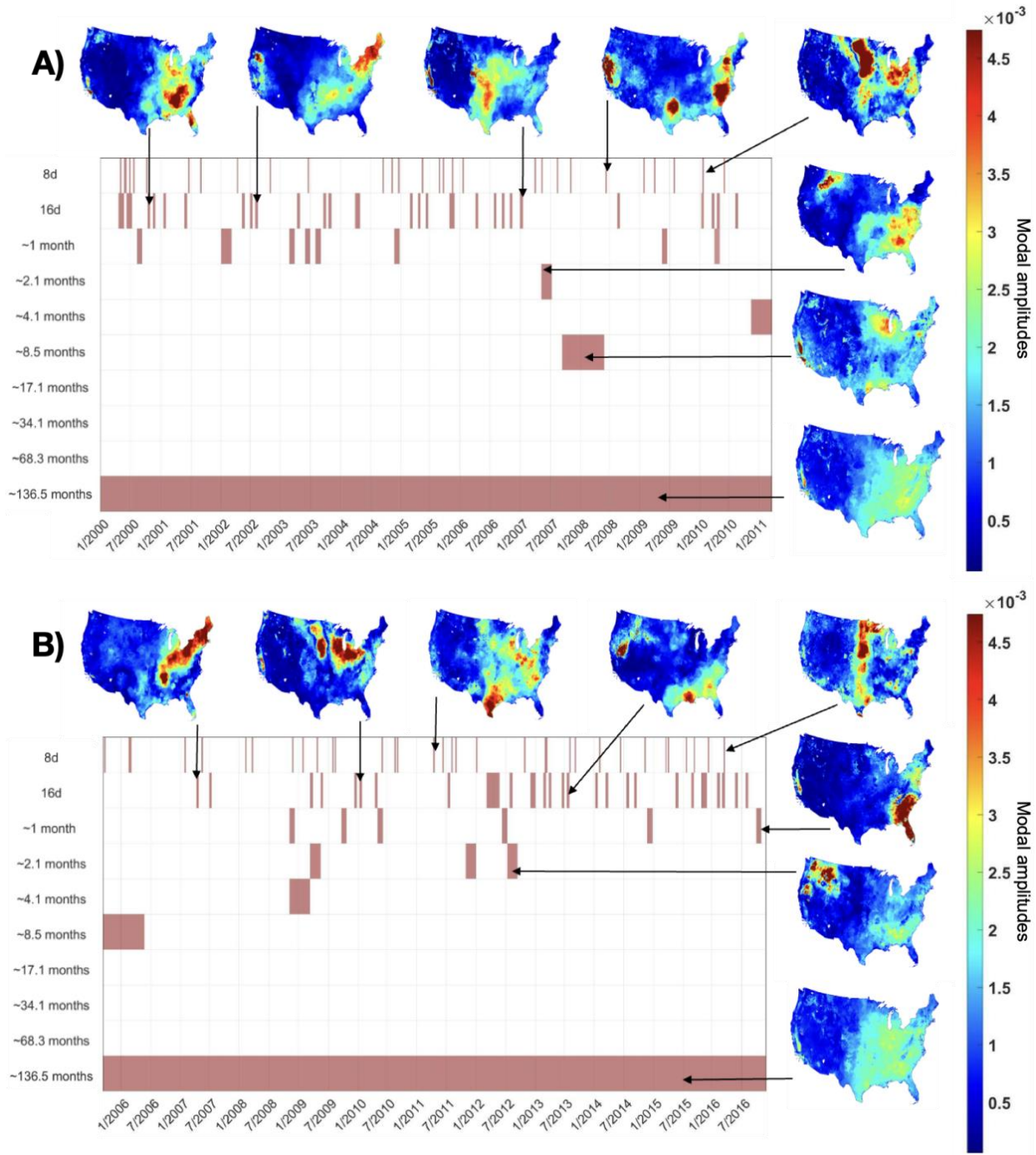

Figure 1. **Maps of mrDMD PM$_{2.5}$ modes from 2000 to 2016**. The mrDMD time windows are for January 2000 to March 2011 (A) and September 2005 to December 2016 (B). The left axis expresses the decomposition level and its related time frequency, such that the bottom row corresponds to the average background mode of PM$_{2.5}$ over 4096 days (~136.5 months), with each successively higher row corresponding to pollution episodes lasting half the number of days of the row below. Colored boxes indicate a mrDMD mode that exhibits significant variability above the background mode; otherwise, the boxes are left blank. No significant pollution modes less than one week are found and thus these rows are removed from the Figure. Modal maps of the background averages and examples of significant pollution episodes are shown in the margins. Arrows point to the time periods of the corresponding mrDMD modal maps. All mrDMD modes are on the same amplitude scale shown to the left of the figure.

Based on Figure 1, we conclude that the mrDMD framework can distinguish high pollution episodes from the mean $PM_{2.5}$ concentrations through time. The mrDMD framework provides a background mode that agrees with Di et al (2019), but also contains modal snapshots of events that stand out from this background.

3.2 Optimal sensor locations

By applying QR pivots to our mrDMD modes, we can identify a set of 1369 optimal sensor locations across the contiguous United States, a number of roughly the same magnitude as the number of EPA sensors. Our method assumes that air pollution spatiotemporal patterns and concentration variability from 2000-2016 are representative of future conditions. We argue that this assumption is reasonable, as discussed below. Figure 2 shows that the mrDMD sensor distribution is more clustered in California and across the West, along the Gulf Coast, and in the Industrial Midwest and Northeast, relative to that of the EPA. The mrDMD sensors thus reflect those regions characterized by higher $PM_{2.5}$ variability, with few sensors in the central United States. These results suggest that the West is a driver of $PM_{2.5}$ variability in the United States due to periodic wildfires, distributed urban sources isolated by rugged terrain, and reduced meteorological mixing. For example, the mrDMD algorithm yields a cluster of sensors in Los Angeles and Salt Lake City as these urban areas are basins surrounded by mountains where the meteorological conditions favor the production and trapping of $PM_{2.5}$ pollution.

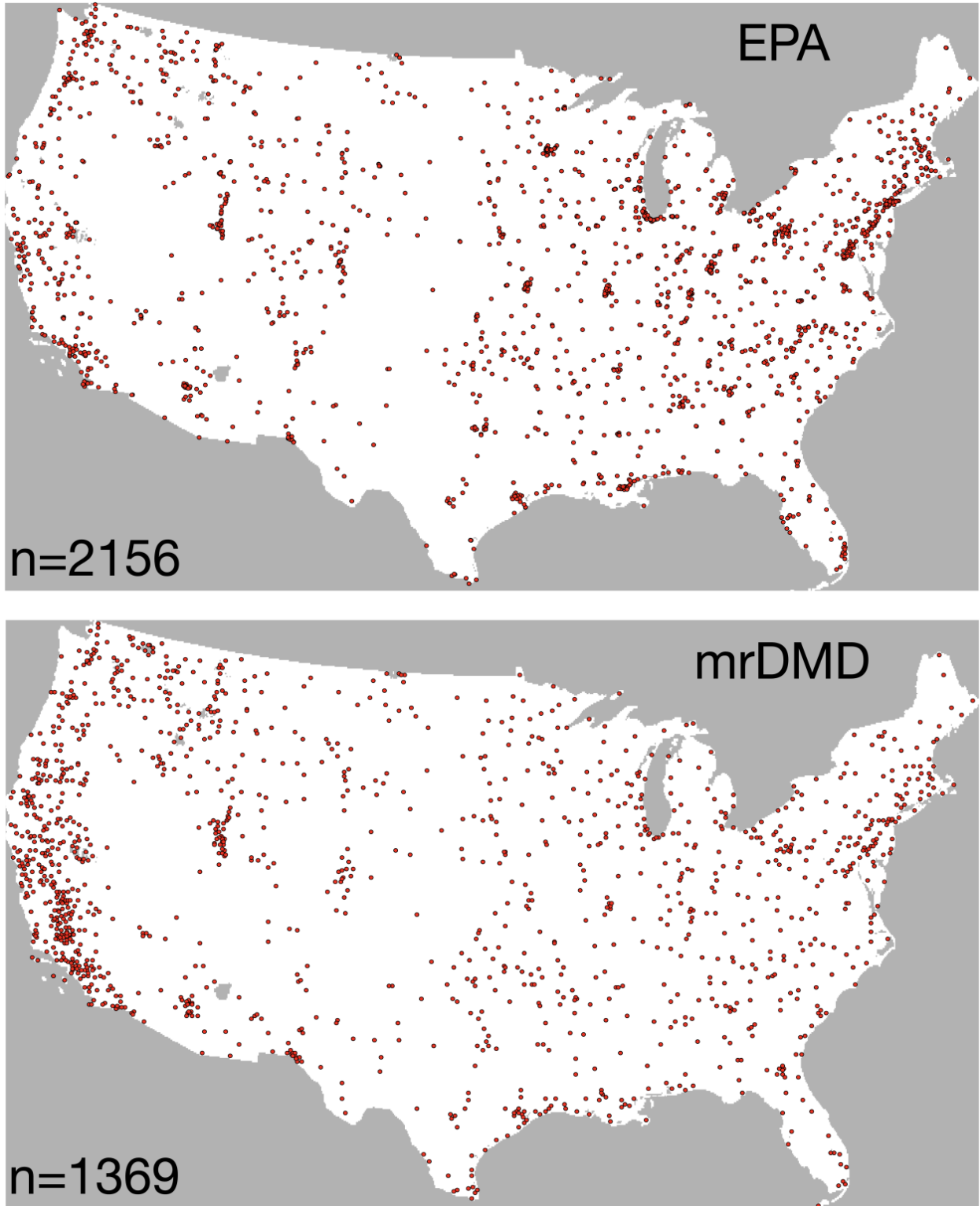

Figure 2. **PM$_{2.5}$ sensor locations**. Distribution of sensor locations in the EPA monitoring network compared to those identified as optimal by the mrDMD algorithm. The EPA sites are shown at original scale while mrDMD locations are projected on a 10 km × 10 km grid. Small

grey patches in the Southwest and Pacific Northwest reflect grid cells with missing or corrupted data that were discarded from this analysis.

The mrDMD method identifies significantly more sensors in the San Joaquin Valley and in Northern California, and the Pacific Northwest than the current EPA sensor network (Figure 3). In particular, the San Joaquin Valley dominates the modal signatures of the mrDMD library (Figure 1) and is characteristic of high $PM_{2.5}$ variability due to large-scale agricultural activities, smoke from wildfires, and frequent temperature inversions. The greater density of mrDMD sensors in northern California and the Pacific Northwest (especially in Idaho and the eastern parts of Oregon and Washington) can be attributed to their sensitivity to smoke $PM_{2.5}$ from wildfires. Indeed, we find that wildfires constitute 60% (46 out of 77) of significant pollution modes in the mrDMD library from September 2005 to December 2016. Also, mrDMD identifies locations along the Gulf Coast (i.e., the eastern shoreline of Texas) that are not covered by the EPA network (Figure S1).

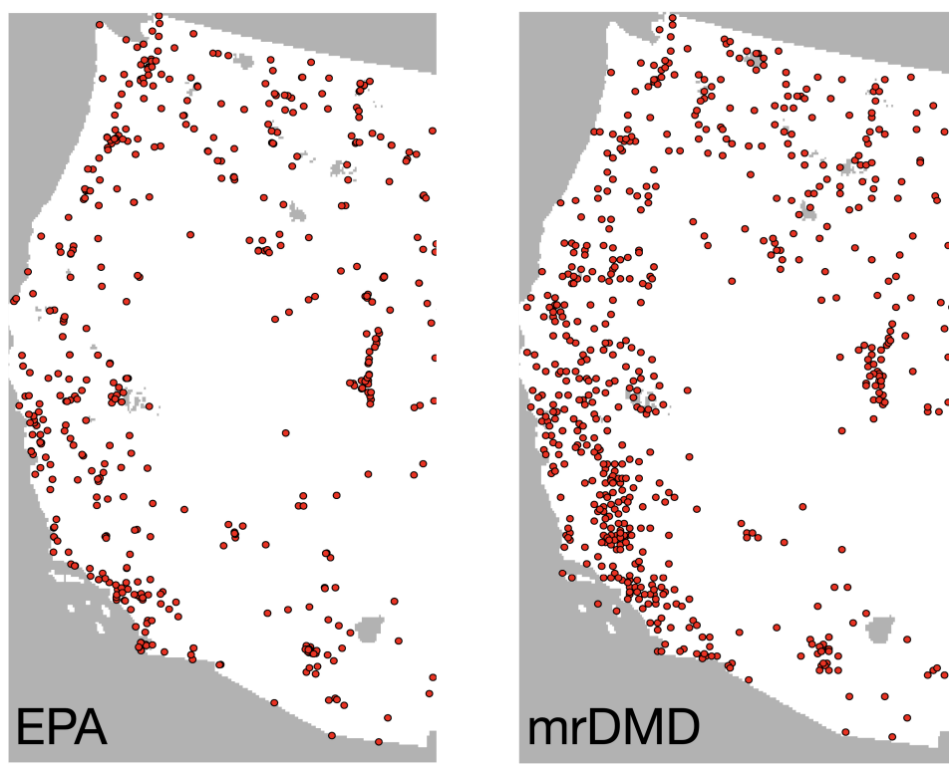

Figure 3. **$PM_{2.5}$ sensor locations in the West**. Same as Figure 2 but zoomed in on the western United States.

Our results suggest that the EPA monitoring network does not optimally capture the spatial patterns and variability of $PM_{2.5}$ pollution in the contiguous United States today. We posit that the current locations of EPA sensors were informed by the legacy of air pollution from the 1980s-1990s when the eastern United States had a higher regional $PM_{2.5}$ background with large variability in $PM_{2.5}$ concentrations arising from interactions between meteorology and emissions from coal-fired power plants, diesel emissions, and industrial activities (Bachmann, 2007). However, with the passage of the Clean Air Act (and subsequent amendments), stricter air quality regulations have reduced these emissions (Bachmann, 2007). Urban air pollution levels in turn declined nationally while at the same time the frequency and magnitude of wildfires in the West have increased significantly (Abatzoglou and Williams, 2016). Our mrDMD sensor locations underscores this shift to the West as 53% of sensors (726 out of 1369) are found west of the 100$^{th}$ meridian (Seager et al., 2017) compared to 32% (694 out of 2156) for the current network of EPA monitors. We emphasize that the current distribution EPA monitors is needed to keep track of adherence to the NAAQS, especially in densely populated regions. We do not advocate for the removal of existing PM$_{2.5}$ sensors, but rather for the deployment of more sensors in the West (especially in the San Joaquin Valley and Pacific Northwest) to capture high-pollution variability.

**Conclusions**

We present a pilot study using a data-driven approach to identify the optimal placement of PM$_{2.5}$ sensors to capture air pollution over the contiguous United States at 10 km × 10 km resolution from 2000 to 2016. While earlier studies are limited in spatial and temporal scope, this is the first national-scale study that diagnoses the optimal placement of PM$_{2.5}$ sensors. Our method employs multiresolution modal decomposition (mrDMD), an approach that incorporates the variability of PM$_{2.5}$ on timescales ranging from one day to over a decade. We identify 72 significant pollution modes from January 2000 to March 2011 and 77 pollution modes from September 2005 to December 2016, ranging in duration from 8 to 256 days. Smoke from wildfire events comprise 38% and 60% of these modes during these respective time frames. Our results underscore the large impact that wildfires have on PM$_{2.5}$ concentrations and variability, especially in the West.

The mrDMD modal library identifies 1369 optimal sensor locations in the western United States, with significantly lower density of sensors in the eastern coast. We show that 53% of mrDMD sensor locations are found west of the 100$^{th}$ meridian compared to only 32% for the current network of EPA monitors. We hypothesize that the current network of EPA monitors was designed to capture PM$_{2.5}$ spatial patterns during the 1980s-1990s when urban air pollution and high background concentrations in the East, coupled with meteorological factors, represented a significant driver of PM$_{2.5}$ variability. However, with reduced emissions from increased air quality regulations, our results suggest a shift to the West as the main driver of PM$_{2.5}$ variability across the contiguous United States. This shift can be traced to periodic wildfires and to distributed urban sources isolated by rugged terrain. We suggest that the mrDMD sensors identified here may help capture high PM$_{2.5}$ episodes not adequately monitored by EPA sensors. We acknowledge, however, that our method assumes that air pollution spatiotemporal patterns and concentration variability from 2000-2016 are representative of future conditions. We argue that this assumption is reasonable, as we expect wildfires to continue to be a significant source of PM$_{2.5}$ and that our optimal sensor locations capture that variability.

More specifically, our mrDMD results recommend adding PM$_{2.5}$ sensors in the San Joaquin Valley in California, northern California, and in the Pacific Northwest (Idaho and the eastern parts of Oregon and Washington). Agricultural activities, temperature inversions, and fires drive pollution variability in the San Joaquin Valley and wildfire smoke drives variability in northern California and the Pacific Northwest. Both regions have large populations of non-white Hispanic agricultural workers who are disproportionately affected by an underserved PM$_{2.5}$ monitoring network (Garcia, 2007; Chandrasekaran, 2021). The disparity in sensor locations is especially apparent in the San Joaquin Valley where the EPA network of monitors is relatively sparse (Figure 3).

Smoke from wildfires represents a significant human health and ecosystem hazard. With the frequency and magnitude of such fires predicted to increase in the future, identifying where to place sensors to track smoke plumes is becoming especially important (David et al., 2021; McClure and Jaffe, 2018). The mrDMD framework skillfully diagnoses where to place air quality sensors that can best monitor smoke plumes, especially those from large fires. We posit that such a sensor distribution may help stakeholders weigh the benefits and hazards of prescribed fires.

Although we identify optimal PM$_{2.5}$ sensor locations on a national scale, our framework may also be impactful for regional and local domains. With the proliferation of low-cost sensors and citizen science efforts, the techniques provided here could help inform high utility areas to place sensors. For example, the original 1 km × 1 km PM$_{2.5}$ dataset by Di et al. (2021) may be applied to investigate sensor placement at the regional and urban scale. Extensions to the DMD framework could incorporate sensor measurements with different signal-to-noise ratios – i.e., high-cost and low-cost sensors (Clark et al., 2020) – or could consider cost-constraining functions that optimize sensor placement based on topography, population density, or metrics related to environmental justice (Clark et al., 2019).

**Acknowledgements**

This work was partly funded by the US Environmental Protection Agency (EPA) grant 83587201. It has not been formally reviewed by the EPA. The views expressed in this document are solely those of the authors and do not necessarily reflect those of the EPA. We thank Yaguang Wei for technical assistance and John Bachmann for valuable discussion.

The DMD algorithm, mrDMD algorithm, and 10 km × 10 km PM$_{2.5}$ dataset from 2000-2016 can be found at Zenodo (https://zenodo.org/record/5809461).

Supplementary information can be found at (https://bit.ly/341o6Fb).

**REFERENCES**

Abatzoglou, J.T., Williams, A.P., 2016. Impact of anthropogenic climate change on wildfire across western US forests. *Proc. Natl. Acad. Sci.* 113, 11770–11775

Askham, T., Kutz, J.N., 2017. Variable projection methods for an optimized dynamic mode decomposition. *ArXiv170402343 Math*

Bachmann, J., 2007. Will the Circle Be Unbroken: A History of the U.S. National Ambient Air Quality Standards. *J. Air Waste Manag. Assoc.* 57, 652–697

Bai, Z., Brunton, S.L., Brunton, B.W., Kutz, J.N., Kaiser, E., Spohn, A., Noack, B.R., 2017. Data-Driven Methods in Fluid Dynamics: Sparse Classification from Experimental Data, in: Pollard, A., Castillo, L., Danaila, L., Glauser, M. (Eds.), Whither Turbulence and Big Data in the 21st Century? Springer International Publishing, Cham, pp. 323–342

Barkjohn, K.K., Gantt, B., Clements, A.L., 2021. Development and application of a United States-wide correction for $PM_{2.5}$ data collected with the PurpleAir sensor. *Atmospheric Meas. Tech.* 14, 4617–4637

Bond, T.C., Bhardwaj, E., Dong, R., Jogani, R., Jung, S., Roden, C., Streets, D.G., Trautmann, N.M., 2007. Historical emissions of black and organic carbon aerosol from energy-related combustion, 1850–2000. *Glob. Biogeochem. Cycles* 21

Brunton, B.W., Johnson, L.A., Ojemann, J.G., Kutz, J.N., 2016. Extracting spatial–temporal coherent patterns in large-scale neural recordings using dynamic mode decomposition. *J. Neurosci. Methods* 258, 1–15

Brunton, S.L., Kutz, J.N., 2019. Data-Driven Science and Engineering: Machine Learning, Dynamical Systems, and Control. Cambridge University Press, Cambridge

Chandrasekaran, P.R., 2021. Remaking "the people": Immigrant farmworkers, environmental justice and the rise of environmental populism in California's San Joaquin Valley. *Journal of Rural Studies* 82, 595–605

Clark, E., Askham, T., Brunton, S.L., Nathan Kutz, J., 2019. Greedy Sensor Placement With Cost Constraints. *IEEE Sens. J.* 19, 2642–2656

Clark, E., Brunton, S.L., Kutz, J.N., 2020. Multi-fidelity sensor selection: Greedy algorithms to place cheap and expensive sensors with cost constraints. *ArXiv200503650 Eess.*

Cusworth, D.H., Mickley, L.J., Sulprizio, M.P., Liu, T., Marlier, M.E., DeFries, R.S., Guttikunda, S.K., Gupta, P., 2018. Quantifying the influence of agricultural fires in northwest India on urban air pollution in Delhi, India. *Environ. Res. Lett.* 13, 044018

David, L.M., Ravishankara, A.R., Brey, S.J., Fischer, E.V., Volckens, J., Kreidenweis, S., 2021. Could the exception become the rule? "Uncontrollable" air pollution events in the U.S. due to wildland fires. *Environ. Res. Lett.* https://doi.org/10.1088/1748-9326/abe1f3

deSouza, P., Kinney, P.L., 2021. On the distribution of low-cost PM2.5 sensors in the US: demographic and air quality associations. J. Expo. Sci. Environ. Epidemiol. 31, 514–524

Dhingra, S., Madda, R.B., Gandomi, A.H., Patan, R., Daneshmand, M., 2019. Internet of Things Mobile–Air Pollution Monitoring System (IoT-Mobair). *IEEE Internet Things J.* 6, 5577–5584

Di, Q., Amini, H., Shi, L., Kloog, I., Silvern, R., Kelly, J., Sabath, M.B., Choirat, C., Koutrakis, P., Lyapustin, A., Wang, Y., Mickley, L.J., Schwartz, J., 2019. An ensemble-based model of PM2.5 concentration across the contiguous United States with high spatiotemporal resolution. *Environ. Int.* 130, 104909

Di, Q., Wei, Y., Shtein, A., Hultquist, C., Xing, X., Amini, H., Shi, L., Kloog, I., Silvern, R., Kelly, J., Sabath, M.B., Choirat, C., Koutrakis, P., Lyapustin, A., Wang, Y., Mickley, L.J., Schwartz, J., 2021. Daily and Annual PM2.5 Concentrations for the Contiguous United States, 1-km Grids, v1 (2000 - 2016). https://doi.org/10.7927/0RVR-4538

Garcia, J., 2007. Mexicans in North Central Washington. San Francisco, CA.: Arcadia Publishing

Goldstein, A.H., Koven, C.D., Heald, C.L., Fung, I.Y., 2009. Biogenic carbon and anthropogenic pollutants combine to form a cooling haze over the southeastern United States. *Proc. Natl. Acad. Sci.* 106, 8835–8840

Heck, L.P., Olkin, J.A., Naghshineh, K., 1998. Transducer Placement for Broadband Active Vibration Control Using a Novel Multidimensional QR Factorization. *J. Vib. Acoust.* 120, 663–670

Johnston, F.H., Henderson, S.B., Chen, Y., Randerson, J.T., Marlier, M., DeFries, R.S., Kinney, P., Bowman, D.M.J.S., Brauer, M., 2012. Estimated Global Mortality Attributable to Smoke from Landscape Fires. *Environ. Health Perspect.* 120, 695–701

Kumar, P., Morawska, L., Martani, C., Biskos, G., Neophytou, M., Di Sabatino, S., Bell, M., Norford, L., Britter, R., 2015. The rise of low-cost sensing for managing air pollution in cities. *Environ. Int.* 75, 199–205

Kutz, J. N. 2013., Data-driven modeling & scientific computation: methods for complex systems & big data. Oxford University

Kutz, J.N., Fu, X., Brunton, S.L., 2016. Multiresolution Dynamic Mode Decomposition. *SIAM J. Appl. Dyn. Syst.* 15, 713–735

Li, Q., Dietrich, F., Bollt, E.M., Kevrekidis, I.G., 2017. Extended dynamic mode decomposition with dictionary learning: a data-driven adaptive spectral decomposition of the Koopman operator. Chaos Interdiscip. *J. Nonlinear Sci.* 27, 103111

Löhner, R., Camelli, F., 2005. Optimal placement of sensors for contaminant detection based on detailed 3D CFD simulations. *Eng. Comput.* 22, 260–273

Manohar, K., Brunton, B.W., Kutz, J.N., Brunton, S.L., 2018. Data-Driven Sparse Sensor Placement for Reconstruction: Demonstrating the Benefits of Exploiting Known Patterns. *IEEE Control Syst. Mag.* 38, 63–86

Manohar, K., Kaiser, E., Brunton, S.L., Kutz, J.N., 2019. Optimized Sampling for Multiscale Dynamics. *Multiscale Model. Simul.* 17, 117–136

McClure, C.D., Jaffe, D.A., 2018. US particulate matter air quality improves except in wildfire-prone areas. *Proc. Natl. Acad. Sci.* 115, 7901–7906

McDuffie, E.E., Martin, R.V., Spadaro, J.V., Burnett, R., Smith, S.J., O'Rourke, P., Hammer, M.S., van Donkelaar, A., Bindle, L., Shah, V., Jaeglé, L., Luo, G., Yu, F., Adeniran, J.A., Lin, J., Brauer, M., 2021. Source sector and fuel contributions to ambient PM2.5 and attributable mortality across multiple spatial scales. *Nat. Commun.* 12, 3594

Mukherjee, R., Diwekar, U.M., Kumar, N., 2020. Real-time optimal spatiotemporal sensor placement for monitoring air pollutants. *Clean Technol. Environ. Policy* 22, 2091–2105

Murray, C.J.L., et al, 2020. Global burden of 87 risk factors in 204 countries and territories, 1990–2019: a systematic analysis for the Global Burden of Disease Study 2019. *The Lancet* 396, 1223–1249

Noack, B.R., Stankiewicz, W., Morzynski, M., Schmid, P.J., 2015. Recursive dynamic mode decomposition of a transient cylinder wake. *ArXiv151106876 Phys.*

Parmar, G., Lakhani, S., Chattopadhyay, M.K., 2017. An IoT based low cost air pollution monitoring system, in: 2017 International Conference on Recent Innovations in Signal Processing and Embedded Systems (RISE). Presented at the 2017 International Conference on Recent Innovations in Signal processing and Embedded Systems (RISE), pp. 524–528

Sashidhar, D., Kutz, J.N., 2021. Bagging, optimized dynamic mode decomposition (BOP-DMD) for robust, stable forecasting with spatial and temporal uncertainty-quantification. *ArXiv210710878 Cs Math*.

Seager, R., Feldman, J., Lis, N., Ting, M., Williams, A. P., Nakamura, J., Liu, H., & Henderson, N, 2017. Whither the 100th meridian? The once and future physical and human geography of America's arid–humid divide. Part II: The meridian moves east. *Earth Interactions*, 22(5), 1–24.

Snyder, E.G., Watkins, T.H., Solomon, P.A., Thoma, E.D., Williams, R.W., Hagler, G.S.W., Shelow, D., Hindin, D.A., Kilaru, V.J., Preuss, P.W., 2013. The Changing Paradigm of Air Pollution Monitoring. *Environ. Sci. Technol.* 47, 11369–11377

Sorensen, A.E., Jordan, R.C., LaDeau, S.L., Biehler, D., Wilson, S., Pitas, J.-H., Leisnham, P.T., 2019. Reflecting on Efforts to Design an Inclusive Citizen Science Project in West Baltimore. *Citiz. Sci. Theory Pract.* 4, 13

Sun, C., Li, V.O.K., Lam, J.C.K., Leslie, I., 2019. Optimal Citizen-Centric Sensor Placement for Air Quality Monitoring: A Case Study of City of Cambridge, the United Kingdom. *IEEE Access* 7, 47390–47400

Tai, A.P.K., Mickley, L.J., Jacob, D.J., 2010. Correlations between fine particulate matter (PM2.5) and meteorological variables in the United States: Implications for the sensitivity of PM2.5 to climate change. *Atmos. Environ.* 44, 3976–3984

Tai, A.P.K., Mickley, L.J., Jacob, D.J., Leibensperger, E.M., Zhang, L., Fisher, J.A., Pye, H.O.T., 2012. Meteorological modes of variability for fine particulate matter ($PM_{2.5}$) air quality in the United States: implications for $PM_{2.5}$ sensitivity to climate change. *Atmospheric Chem. Phys.* 12, 3131–3145

Tissot, G., Cordier, L., Benard, N., Noack, B.R., 2014. Model reduction using Dynamic Mode Decomposition. Comptes Rendus Mécanique, Flow separation control 342, 410–416

Toma, C., Alexandru, A., Popa, M., Zamfiroiu, A., 2019. IoT Solution for Smart Cities' Pollution Monitoring and the Security Challenges. *Sensors* 19, 3401

US EPA, 2021. Clean Air Plans; California; San Joaquin Valley Moderate Area Plan and Reclassification as Serious Nonattainment for the 2012 PM2.5. [WWW Document]. URL https://www.federalregister.gov/documents/2021/09/01/2021-18764/clean-air-plans-california-san-joaquin-valley-moderate-area-plan-and-reclassification-as-serious. (accessed 11.15.21).

US EPA, 2020. Ambient Air Monitoring Quality Assurance Guidance Documents [WWW Document]. URL https://www.epa.gov/amtic/ambient-air-monitoring-quality-assurance-guidance-documents (accessed 9.21.21).

US EPA, 2013. Air Quality System (AQS) [WWW Document]. URL https://www.epa.gov/aqs (accessed 9.21.21).

Waeytens, J., Sadr, S., 2018. Computer-aided placement of air quality sensors using adjoint framework and sensor features to localize indoor source emission. *Build. Environ.* 144, 184–193

Yildirim, B., Chryssostomidis, C., Karniadakis, G.E., 2009. Efficient sensor placement for ocean measurements using low-dimensional concepts.